\documentstyle[12pt,amssymb]{article}
\newcommand{\beq}[1]{\begin{equation}\label{#1}}
\newcommand{\eeq}{\end{equation}}
\newcommand{\bear}[1]{\begin{eqnarray}\label{#1}}
\newcommand{\ear}{\end{eqnarray}}
\newcommand{\nn}{\nonumber}

\newcommand{\rf}[1]{(\ref{#1})}
\newcommand{\nl}{ {\hfill \break} }
\newcommand{\np}{ {\newpage } }

\newcommand{\iso}{ {\cong } }

\newcommand{\Iff}{ {\Leftrightarrow } }

\newcommand{\imp}{\ {\Rightarrow }\ }

\newcommand{\partl}{ {\partial } }
\newcommand{\cl}{ { \mathrm{cl} } }
\newcommand{\clc}{ { \mathrm{cl}_c} }

\newcommand{\N}{ \mbox{\rm I$\!$N} }
\newcommand{\R}{ \mbox{\rm I$\!$R} }
\newcommand{\Z}{ \mbox{$\mathbb Z$} }
\newcommand{\Diff}{ \mbox{\rm Diff} }
\newcommand{\Out}{ \mbox{\rm Out} }
\newcommand{\Inn}{ \mbox{\rm Inn} }
\newcommand{\Aut}{ \mbox{\rm Aut} }

\newcommand{\tr}{ \mbox{\rm tr} }

\newcommand{\SO}{ \mbox{\rm SO} }

\newcommand{\Int}{ {\mathrm int} }
\newcommand{\Img}{ {\mathrm im} }

\newcommand{\Cinf}{ {\mathcal{C}}^\infty }
\newcommand{\Cyl}{ \mbox{\rm Cyl} }
\newcommand{\Der}{ \mbox{\rm Der}(\Cyl) }

\renewcommand{\setminus}{-}

\def\Journal#1#2#3#4{{#1} {\bf #2}, #3 (#4)}

\def\ATMP{\em Adv. Theor. Math. Phys.}
\def\CMP{\em Commun. Math. Phys.}

\def\IJTP{\em Int. J. Theor. Phys.}

\def\be{\begin{equation}}
\def\ee{\end{equation}}
\def\bea{\begin{eqnarray}}
\def\eea{\end{eqnarray}}
\DeclareFontFamily{U}{rsfs}{}         % Formal Script            %
\DeclareFontShape{U}{rsfs}{m}{n}{<5> rsfs5 <6><7> rsfs7          %
  <8><9><10><10.95><12><14.4><17.28><20.74><24.88> rsfs10}{}     %
\DeclareMathAlphabet{\mathfs}{U}{rsfs}{m}{n}                     %
\newcommand{\mfs}[1]{\mathfs {#1}}                               %
\newcommand{\sC}{ {\mfs C}}
\newcommand{\sK}{ {\mfs K}}
\newcommand{\sF}{ {\mfs F}}
\newcommand{\sP}{ {\mfs P}}
\newcommand{\sT}{ {\mfs T}}
\newcommand{\sE}{ {\mfs E}}
\newcommand{\sH}{ {\mfs H}}

\newcommand{\olcirc}[1]{\vbox{\mathsurround=0pt                %
  \skip1=0pt plus 1 fil  \skip3=0pt plus 3fil                    %
  \ialign{##\crcr$\hskip\skip3\scriptstyle\circ\hskip\skip1$\crcr%
    \noalign{\kern1pt\nointerlineskip}$\displaystyle{#1}$\crcr}} %
  \vphantom{#1}}                                                 %
\newcommand{\oscirc}[1]{\vbox{\mathsurround=0pt                %
  \skip1=0pt plus 2 fil  \skip3=0pt plus 2fil                    %
  \ialign{##\crcr$\hskip\skip3\scriptstyle\circ\hskip\skip1$\crcr%
    \noalign{\kern1pt\nointerlineskip}$\displaystyle{#1}$\crcr}} %
  \vphantom{#1}}                                                 %
\begin{document}
%\title{\bf\large On causal nets of C$^*$-algebras
%in quantum gravity}
%\title{\bf\large Causal nets of C$^*$-algebras in
%canonical quantum gravity}
%\title{\bf\large On the Haag-Kastler QFT structure of
%canonical quantum gravity}
%\title{\bf\large Algebraic QFT structures for quantum gravity }
\title{\bf\large Algebraic Quantum Theory on Manifolds:
\\
A Haag-Kastler Setting for Quantum Geometry}

%off%
\author
{Martin Rainer\\
%off%\address{
%\begin{center}{
Center for Gravitational Physics and Geometry, 104 Davey Laboratory,\\
The Penn State University, University Park, PA 16802-6300, USA \\
and\\
Mathematische Physik I, Mathematisches Institut,\\
Universit\"at Potsdam, PF 601553, D-14415 Potsdam, Germany
}
%}\end{center}
\date{May 31, 1999}

%%%%%%%%%%%%%%%%%%%%%%%%%%%%%%%%%%%%%%%%%%%%%%%%%%%%%%%%%%%%%%
%%%%%%%%%%%%%%%%%%%%%%%%%%%%%%%%%%%%%%%%%%%%%%%%%%%%%%%%%%%%%%
\maketitle
%\vspace*{-8.5cm}
%\centerline{\mbox{\hspace*{10cm}  }}
%\centerline{\mbox{\hspace*{10cm}
%}}
%\centerline{\mbox{\hspace*{10cm} }}
%off%\abstracts{
\centerline{\bf Abstract}
\vspace*{0.9cm}
{
Motivated by the invariance of current
representations of quantum gravity under diffeomorphisms
much more general than isometries,
the Haag-Kastler setting is extended to
manifolds without metric background structure.
First, the causal structure on a differentiable
manifold $M$ of arbitrary dimension $(d+1>2)$ can be  defined in purely topological terms,
via cones (C-causality).
Then, the general structure
of a net of $C^*$-algebras on a manifold $M$
and its causal properties
required for an algebraic quantum field theory
can be described as an extension of the
Haag-Kastler axiomatic framework.

An important application
is given with quantum geometry on a spatial slice $\Sigma$
within the causally exterior region of a
topological horizon $\sH$,
resulting in a net of Weyl algebras
for states with
an infinite number of intersection
points of edges and
transversal $(d-1)$-faces
within any neighbourhood of the spatial boundary
$\sH{\rm }\cap\Sigma\iso S^2$.
\np
%%%%%%%%%%%%%%%%%%%%%%%%%%%%%%%%%%%%%%%%%%%%%%%%%%%%%%%%%%%%%%%%%%%%%%%%%%
\section{Introduction} %1
%%%%%%%%%%%%%%%%%%%%%%%%%%%%%%%%%%%%%%%%%%%%%%%%%%%%%%%%%%%%%%%%%%%%%%%%%%
While classical general relativity usually
employs a Lorentzian spacetime structure,
the most successful approaches for quantum gravity,
such as the canonical quantization of the connection representation,
the loop representation or the spin network representation,
and topological quantum field theory, BF-theories and spin foams
are  invariant under diffeomorphism much more general
than isometries.
Currently, most of these approaches are mainly proposals for the
quantum theory of some free geometry.
Although in particular situations
also coupling to matter has been studied,
at present in spacetime dimension $3+1$
no complete theory for quantum geometry
coupled to quantum matter is available.
This situation is remarkably paralleled by
the algebraic framework of quantum field theory (QFT)
on a curved spacetime of dimension $3+1$, which
is most successful in describing the structure of
free or asymptotically free quantum field theories.
However,
the standard Haag-Kastler quantum field theories
over Minkowski space or curved asymptotically free spacetimes
is not  invariant under diffeomorphisms
unless these are isometries.

Recently a diffeomorphism invariant extension
of the axiomatic Osterwalder-Schrader framework \cite{OS}
of constructive Euclidean QFT \cite{GJ} has been given
by \cite{AMMT}. There the Osterwalder-Schrader reconstruction
of Euclidean QFT was generalized. However there is no obvious
generalization of those analyticity properties which
relate the Schwinger functions and Wightman functions
of the standard Euclidean and Minkowskian QFT.
At present there is no notion of
a generalized "Wick rotation"
which would allow us to use the results of  \cite{AMMT}
in order to infer properties of an appropriate
diffeomorphism invariant generalization
of the algebraic Haag-Kastler framework for Lorentzian QFT.

Recently the algebraic framework was used sucessful
in providing a clear proof \cite{Reh} of the holographic hypothesis
(Maldacena conjecture \cite{Mal,Wit}). While there is no direct point
transformation between the bulk degrees of freedom
in anti-de Sitter (AdS) space and those of the boundary CFT,
the relation becomes clear for the corresponding
algebras localized in wedge regions on AdS space
and in double cones on its boundary.

Therefore it is the goal of the present investigations
to extend the framework of
algebraic quantum field theory (AQFT) and Haag-Kastler axioms
such that  it becomes also applicable
to theories
which are invariant
under a larger class
of diffeomorphisms, such as quantum gravity.
Within the AQFT setting the question
rises what are the general
classes of diffeomorphisms compatible with a given
algebraic structure.

Earlier attempts \cite{Ra97,RaSa,Ra1} towards a diffeomorphism invariant
algebraic setting for quantum field theory
mainly generalized those axioms that use only
a topological structure on the manifold
already in the usual setting,
such as isotony and  covariance.
However the causality axiom could only be formulated
in a very rudimentary sense, namely by
generalizing Haag duality just on the boundary
of the net. That procedure introduced quite
strange features on the net of $*$-algebras.
In particular it implied the existence of
an Abelian center in the localized algebras
which was then associated with the minimal (interior) boundary
\cite{Ra97}.
The existence of such an Abelian center implied
in particular that the algebras could not be
usual CCR Weyl algebras,
since these would have to be simple if the
symplectic form was nondegenerate.

%and the existence of a
%cyclic invariant vector in the GNS Hilbert space
%of a given state.

Nevertheless, presently it is known that
the algebraic structure of
a free quantum field theory on Minkowski space
or an asymptotically free field theory on a
(usually globally hyperbolic) curved space-time
is  encoded in a \emph{ causal} net
of C$^*$-algebras.
In particular on Minkowski space,
there exists a strong correspondence
between particular causal sets of Minkowski space
and localized C$^*$-algebras of the net.
There are particular causal sets
which form a topological basis
of Minkowski space,
namely the bounded double cones.
Moreover, for a net of subalgebras of a \emph{Weyl} algebra,
it is possible \cite{Ba1} to work with a flexible notion of
causality rather than with a rigidly given one,
and in principle the net together with
its underlying manifold might
be reconstructed from the relation among the
localized C$^*$-algebras only \cite{Ba2}.

This motivates the present approach where we present
a generalization of this causal net structure
in an a priori background independent
("diffeomorphism invariant") manner.
Here the net has to be background independent,
but still compatible
with a  (metric independent)
notion of causality.
In order to achieve this, we  have to abstract
an appropriate {\em topological} notion
of causality.
Appropriate definitions for such a notion
were given recently in \cite{Ra99}.
Those diffeomorphism which preserve
such topological causal structure,
should naturally also leave invariant
the algebraic structure of the net.
A causal topology on a $d+1$-dimensional
manifold then provides a topological notion of
a causal complement on any set.
The next step is then to find a natural implementation
of the correspondence between
causal sets  on a differentiable manifold
and C$^*$-algebras localized on these sets.
This amounts to define a causal net of  C$^*$-algebras
on causal differentiable manifolds.
%\nl

The first condition
which a causal net of C$^*$-algebras
over a causal topological space
should naturally satisfy is
that C$^*$-algebras over causally disjoint sets
mutually commute.
Note that e.g. in a $d+1$-dimensional black hole spacetime,
%the interior of a black or white hole horizon is
%the causal complement of its exterior,
%while on
the intersection of a spatial slice with
the horizon is a $d-1$-dimensional sphere $S^{d-1}$.
The latter may be viewed as the boundary of a
minimal $d$-dimensional open set ${O}_{\rm min}$
contained in any
larger $d$-dimensional open set  ${O}_{\rm max}$
within a spatial slice
which extends through all of the region exterior
of the horizon up to spatial infinity $i^0$.
In \cite{Ra97} a generalization of Haag duality
was implemented algebraically,
by demanding that the commutant of the (asymptotic) global
algebra equals a minimal Abelian center algebra
located over the minimal set.

The possible results of a concrete observation
are encoded in a corresponding state
on the causal net of C$^*$-algebras.
Particularly convenient states for quantum geometry
are the spin-network states.
A state introduces additional structure
which may serve to distinguish gauge invariances
from more genuine symmetries (like unitarity of the dynamics).
Given a causal foliation
of  spatial slices exterior to a topological horizon
(via causal boundaries), those diffeomorphisms which
leave invariant the causal foliation are purely
gauge.

In quantum geometry spin-network states
are given via an embedding of a closed graph into
a particular slice of the foliation.
Then there exists diffeomorphisms which keep
the set of all slices invariant
but change the foliation monotonously,
preserving the natural order of the slices.
These are topological dilations.
If there was no state two foliations related by such
a change should
be considered as equivalent.
However an embedded graph can eventually
detect a monotonous change in the foliation
by a change in the origial relations between
the slices and the graph, which are given by the
intersections of the edges of the graph with
the slices of the foliation.
The relations between an embedded graph
and a foliation are encoded topologically
in the
intersections of edges with slices of a foliations.
Change of this intersection topology by dilating one slice onto
another can result in changes of the C$^*$-algebra.
Therefore dilation diffeomorphisms can not be gauge here,
but rather should correspond to outer (i.e. non-trivially
represented) isomorphisms on the algebras.

%%%%%%%%%%%%%%%%%%%%%%%%%%%%%%%%%%%%%%%%%%%%%%%%%%%%%%%%%%%%%%%%%%%%%%%%%
\section{Cone causality on  differentiable $d+1$-manifolds} %3
%\setcounter{equation}{0}
%%%%%%%%%%%%%%%%%%%%%%%%%%%%%%%%%%%%%%%%%%%%%%%%%%%%%%%%%%%%%%%%%%%%%%%%%
Let us now define the notion of a causal topology for general
differentiable $d+1$-manifold $M$ within
any differentiability category which
unless specified otherwise should not be larger than $C^1$
and for convenience may be taken $C^\infty$.
(With minor modifications an extension to the $C^0$ category
is possible too \cite{Ra99}. However for the present
purpose the differentiable setting is most convenient.)
Let
\beq{scone}
\sC:=\{x\in\R^{d+1}:x_0^2=(x-x_0 e_0)^2\} ,
\sC^+:=\{x\in \sC:x_0\geq 0\} ,
\sC^- :=\{x\in \sC:x_0\leq 0\}
\eeq
be the standard (unbounded double) light cone,
and the forward and backward subcones in $\R^{d+1}$, respectively.
The standard open interior and exterior of $\sC$ is defined as
\beq{scone2}
\sT:=\{x\in\R^{d+1}: x_0^2>(x - x_0 e_0)^2\} ,
\sE:=\{x\in\R^{d+1}: x_0^2<(x - x_0 e_0)^2\} .
\eeq
A \emph{manifold thickening} with thickness $m>0$ is given as
\beq{thickcone}
\sC^m:=\{x\in\R^{d+1}:|x_0^2-(x - x_0 e_0)^2|<m^2\} ,
\eeq
The characteristic topological data of the standard cone
is encoded in the topological relations of all its
manifold subspaces
(which includes in particular also the singular vertex $O$)
and among each other.

\textbf{Definition 1:}
Let $M$ be a differentiable manifold.
A  \emph{(null) cone} at $p\in \Int M$ is
the diffeomorphic image $\sC_p:=\phi_p \sC$
of a diffeomorphism of topological spaces
$\phi_p: \sC \to \sC_p \subset  M $
with $\phi_p(0)=p$, such that
\\
(i) every (differentiable) submanifold $N\subset \sC$
is mapped  diffeomorphically on
a submanifold  $\phi_p(N)\subset M$,
\\
(ii) for any two submanifolds $N_1,N_2\subset \sC$
there exist diffeomorphisms
$ \phi_p(N_1) \cap \phi_p(N_2) \cong N_1 \cap N_2$
and $\phi_p(N_1) \cup \phi_p(N_2)\cong N_1 \cup N_2$,
\\
and (iii) for any two differentiable
curves $c_1,c_2:]-\epsilon,\epsilon[\to\sC$
with $c_1(0)=c_2(0)=p$ it holds
$T_0 c_1=T_0 c_2
%%% corrected 2000
\Leftrightarrow
%%%
T_p(\phi_p\circ c_1)|_{ ]-\epsilon,\epsilon[ }
=T_p (\phi_p \circ c_2)|_{ ]-\epsilon,\epsilon[ }$.

Condition (iii) says that
the well defined notion of transversality of intersections
at the vertex is preserved by $\phi_p$.

\textbf{Definition 2:}
An \emph{(ultralocal) cone structure} on $M$ is
an assignment
$\Int M\ni p\to\sC_p$ of a cone
at every $p\in\Int M$.

Note that, although $\sC_p=\phi_p(\sC)$, $\sT$ and $\sE$ here need
not be diffeomorphic to $\phi_p(\sT)$ and $\phi_p(\sE)$ respectively.
A cone structure on $M$ can in general
be rather wild
with cones at different points totally unrelated
unless we impose a topological connection
between the cones at different points.
The cone structure can be tamed
by the requirement of differentability
of the family $\{p\to\sC_p\}$.

\textbf{Definition 3:} Let $M$ be a differentiable manifold.
A \emph{weak ($\sC$) local cone (LC) structure}   on $ M$ is
a  cone structure which is differentiable
i.e. $\{p\to\sC_p\}$ is a differentiable family.

A weak LC structure at each point $p\in\Int  M$
defines
a characteristic topological space $\sC_p$ of codimension $1$
which is Hausdorff everywhere but at $p$.
In particular $\sC_p$ does not contain any
open $U\ni p$ from the manifold topology of $M$.
This can be  improved by resolving the cone.

\textbf{Definition 4:}
Let $M$ be a differentiable manifold.
A  \emph{(manifold) thickened cone} of thickness $m>0$ at $p\in\Int  M$
is the  diffeomorphic image $\sC^{m}_{p}:=\phi_p \sC^m$
of a  diffeomorphism of manifolds
$\phi_p: \sC \to \sC_p \subset  M $
with $\phi_p(0)=p$.

Note that due to the manifold property
a thickened cone is much more simple than a cone itself.

\textbf{Definition 5:}
A \emph{thickened cone structure} on $M$ is
an assignment
$\Int M\ni p\mapsto\sC^{m(p)}_{p}$ of a
thickened cone
at every $p\in\Int  M$.

Note that in general the thickness $m$
can vary from point to point in $M$.
However it is natural to require $m:M\to\R_+$
to be differentiable.
\\
\textbf{Definition 6:} Let $M$ be a differentiable manifold.
A \emph{strong  ($\sC^m$) LC structure}  on $ M$ is
a  differentiable family
of  diffeomorphisms
$\phi_p: \sC^m \to \sC^{m(p)}_{p} \subset M $
with $\phi_p(0)=p$ and
such that the thickness $m$ is a differentiable function on $M$.

Note that a strong LC structure is still much more flexible than a conformal structure.
%off the vertex
For any $q\neq p$
the tangent directions given by $T_q\sC_p$ need a priori
not be related
to tangent directions of null curves of $g$,
since the cone (or its thickening) at $p$ is
in general unrelated to that at $q$.

\textbf{Theorem 1:}
Let $M$ carry a strong LC structure.
At any $p\in \Int M$ there exists an open $U\ni p$
such that:
\\
For $d:=\dim M-1>0$
it is  $|\Pi_{0}(\sT_p|_U)|=2$ and
$\Pi_{d-1}(\sE_p|_U)=\Pi_{d-1}(S^{d-1})$,
\\
for $d>1$ it is
$\Pi_{d-1}(\sT_p|_U)=0$ and $|\Pi_{0}(\sE_p|_U)|=1$,
\\
for $d=1$ it is
$\Pi_{d-1}(\sT_p|_U)=\Pi_{d-1}(\sE_p|_U)=\Pi_{0}(S^{0})$, i.e.
$|\Pi_{0}(\sT_p|_U)|=|\Pi_{0}(\sE_p|_U)|=2$,
\\
and in dimension $d=0$ it is
$\sT_p=\sE_p=\emptyset$.
\\
\emph{Proof:}
For all $p\in \Int M$ the strong LC structure
provides a thickened cone $\sC^{m(p)}_p$.
Since $m(p)>0$, $\sC^{m(p)}_p$ contains always a neighborhood $U\ni p$
diffeomorphic to a neighborhood $\phi^{-1}_p(U)\ni 0$
of the standard cone which in any dimension has the
desired properties.
\hfill \mbox{$\square$}

At any interior
point $p\in\Int  M$
the open exterior $\sE_p$ and
the open interior $\sT_p$ of the cone $\sC_p$ are locally
 topologically distinguishable for $d>1$,
indistinguishable for $d=1$, and empty for $d=0$.
With a strong LC structure,
$\sT_p|_U \neq\sE_p|_U \forall U\ni p$ $\iff$ $d+1>2$,
whence locally in any neighborhood $U\ni p$
the interior and exterior of $\sC_p\cap U$ at $p$ in $U$
has an intrinsic invariant meaning.
$\sC_p|_U$ divides $U\setminus \sC_p|_U$ in three
open submanifolds,
a non-contractable exterior $\sE_p|_U$,
plus two contractable connected components of
$\sT_p=:\sF_p|_U\cup\sP_p|_U$, the local future
$\sF_p|_U$ and the local past $\sP_p|_U$ with
$\partl \sF_p|_U=\sC^+_p|_U$ where $\sC^+_p:=(\phi_p \sC^+)$
and $\partl \sP_p|_U=\sC^-_p|_U$ where $\sC^-_p:=\phi_p \sC^-$
respectively.
This rises also the question if and how $\sF_p$
and $\sP_p$ or their local restriction to $U\ni p$
can be distinguished.
This problem is solved by a topological ${\Z}_2$ connection.
%(see also Section III below).

Let $M$ be differentiable
and $\tau$ be any vector field  $M\to TM$
such that at any $p\in\Int  M$ its orientation
agrees with that of $\phi_p(a)$.
Such a orientation vector field comes naturally along with a
$\Z_2$-connection on $M$
which allows to compare the orientation $\tau(p)$ at different $p\in\Int  M$.
Given a strong LC structure on $M$,
the $\Z_2$-connection is in fact provided via continuity of $p\mapsto T_p\phi_p(a)$.
Then $\tau$ is tangent to an integral curve segment
through $p$ from  $\sP_p$ to $\sF_p$.
In particular,
$\sF_p$ and $\sP_p$ are distinguished from
each other by a consistent $\tau$-orientation on $M$.

In order to obtain a causal structure
it remains to identify natural consistency conditions
for the intersections of cones of different points.
In order to define topologically
timelike, lightlike, and spacelike relations,
and a reasonable causal complement,
we introduce the following causal consistency
conditions on cones.
\\
\textbf{ Definition 7:}
$M$
%\cong\R\times\Sigma$
is \emph{(locally) cone causal} or \emph{C-causal} in an open region $U$,
if it %is (locally) precausal
carries a (weak or strong) LC structure
and in  $U$
the following relations between different cones in $\Int M$ hold:
\\
(1) For $p\neq q$ one and only one of the following is true:
\\
(i) $q$ and $p$ are causally \emph{timelike} related, $q\ll p$ $:\Iff$
$ q\in \sF_p$ $\wedge$ $p \in \sP_q $
(or $p \ll  q$)
\\
(ii) $q$ and $p$ are causally \emph{lightlike} related, $q\lhd  p$  $:\Iff$
$ q\in \sC^+_p\setminus \{p\}$ $\wedge$ $p \in \sC^-_q\setminus \{q\} $
(or $p \lhd  q$),
\\
(iii) $q$ and $p$ are causally unrelated,
i.e. relatively \emph{spacelike} to each other, $q\bowtie  p$ $:\Iff$
$ q\in \sE_p$ $\wedge$ $p \in \sE_q $.
\\
(2) Other cases (in particular non symmetric ones)
do not occur.
\\
$M$ is {\it locally C-causal}, if it is C-causal in any  region $U\subset M$.
$M$ is {\it C-causal} if conditions (1) and (2) hold $\forall p\in \sC$.

Let $M$ be C-causal in $U$.
Then, $q \ll  p$ $\Iff$ $\exists r: q\in \sP_r \wedge p\in \sF_r $,
and $q \lhd  p$ $\Iff$ $\exists r: q\in \sC^+_r \wedge p\in \sC^-_r $.

If an open curve $\R\ni s\mapsto c(s)$ or a
closed curve $S^1\ni s\mapsto c(s)$ is embedded in $M$, then in particular
its image is $\Img(c)\equiv c(\R)\cong \R$ or
$\Img(c)\equiv c(S^1)\cong S^1$ respectively,
whence it is free of selfintersections.
Such a curve is called \emph{spacelike}
$:\Iff$ $\forall p\equiv c(s)\in \Img(c) \exists \epsilon :
c|_{]s-\epsilon,s+\epsilon[\setminus\{0\}}\in\sE_{c(s)}$,
and \emph{timelike}
$:\Iff$ $\forall p\equiv c(s)\in \Img(c) \exists \epsilon :
c|_{]s-\epsilon,s+\epsilon[\setminus\{0\}}\in\sT_{c(s)}$.

Note that C-causality of $M$ forbids a multiple refolding intersection
topology for any two cones with different vertices.
In particular along any timelike curve
the future/past cones do not intersect,
because otherwise there would exist points which are simultaneously
timelike and lightlike related.
Continuity then implies that future/past cones in fact foliate the part of $M$
which they cover. Hence, if there exists  a fibration
$\R\hookrightarrow\Int M\twoheadrightarrow \Sigma$,
then C-causality implies in particular that the future/past cones foliate
on any fiber.
Therefore C-causality
%for any open region $U\subset M$
allows a reasonable definition of  a causal complement.

{\bf Definition 8:}
A {\em causal complement} in a set $U$ is a map
$P(U)\ni S\mapsto S^{\perp}\in P(U)$ such that

(i) $S \subseteq S^{\perp\perp}$

(ii) $(\bigcup_j S_j)^\perp=\bigcap_j (S_j)^\perp$

(iii) $S\cap S^\perp=\emptyset$

\textbf{Example 1:}
For any open set $S$ in a C-causal manifold $M$
the \emph{causal complement}
is defined as
\beq{cc}
S^\perp:=\bigcap_{p\in \cl S}{ \sE_p} ,
\eeq
where $\cl S$ denotes the closure in the topology
induced from the manifold.
Although the causal complement is always  open,
it will in general  not be a contractable region even if $S$ itself
is so.

Assume $p$ and $q$ are timelike related, $p\in\sP_q$ and $q\in \sF_p$.
$\sK^q_p:=\sF_p\cap\sP_q$ is the bounded open double cone between $p$ and $q$.
Given any bounded open $\sK$ such that $\exists p,q\in M: \sK=\sF_p\cap\sP_q$,
we set $i^+(\sK):=\{q\}$,  $i^-(\sK):=\{p\}$, and $i^0(\sK):=\sC^+_p\cap\sC^-_q$.
For any $\sK^q_p\subset M$ let
$\clc(\sK^q_p):=
\sK^q_p\cup\left[
( \partial \sF_p\cup\partial \sP_q)
\cap \partial  \sK^q_p
%\cap ( \sC_p\cup \sC_q)
\right]$
be its {\it causal closure}.

Since C-causality prohibits relative refolding of cones, it also ensures that
$(\sK^q_p)^{\perp\perp}=\sK^q_p$, i.e. double cones are closed under
$(\cdot)^{\perp\perp}$.

%%%%%%%%%%%%%%%%%%%%%%%%%%%%%%%%%%%%%%%%%%%%%%%%%%%%%%%%%%%%%%%%%%%%%%%%%
\section{Causal index sets and diffeomorphisms} %3
%\setcounter{equation}{0}
%%%%%%%%%%%%%%%%%%%%%%%%%%%%%%%%%%%%%%%%%%%%%%%%%%%%%%%%%%%%%%%%%%%%%%%%%
Let us now define the index sets which will be used in
our nets of algebras.
The natural numbers $\N$ are the most common index set for any
countable set on which they induce then a canonical order relation.
However, in the following we consider more general index sets
which need not be countable.

{\bf Definition 9:}
A {\em net index set} is an index set $I$
(i) with a partial order $\leq$,
(ii) with a sequence of $K_i\in I$, ${i\in\N}$,
such that $\forall K\in I \exists j\in\N$: $K\leq K_j$,
and (iii) such that each bounded $J\subset I$
has a unique supremum $\sup J\in I$.
\\
Remark 1: If $I$ is totally ordered (iii) is satisfied
trivially.
\\
Remark 2: By (ii) a net index set is infinite
unless $\exists j\in\N$: $K_i=K_j$ $\forall i\geq j$.
%Note that any finite lattice $L$ can always be labelled
%by indices from a net index set.
%Consider $L\ni m_1,\ldots,m_k$ such that
%$l\in L \Imp \exists m_i\geq l$. The finite sequence
%$1,\ldots,k$ can then be trivially extended

{\bf Definition 10:}
A {\em causal disjointness relation} in a net index set $I$
is a symmetric relation $\perp$ such that

(i) $K_1\perp K_0$ $\wedge$ $K_2 < K_1$  $\imp$ $K_2\perp K_0$,

(ii) for any bounded $J\subset I$:
$K_0\perp K$ $\forall K\in J$ $\imp$ $K_0\perp \sup J$,

(iii) $\forall K_1\in I$ $\exists K_2\in I$: $K_1\perp K_2$.

%{\bf Definition:}
A {\em causal index set} $(I,\perp)$ is a
net index set with a causal disjointness relation $\perp$.
%such that ....
%there is a sequence of $K_i\in I$, ${i\in\N}$,
%such that $\forall K\in I \exists i_0\in\N$: $K\leq K_i$.

{\bf Definition 11:}
Let $M$ be infinite with causal complement $\perp$.
$M$ is {\em $\perp$-nontrivially inductively covered},
iff $\exists$ sequence of nonempty $K_i\subset M$, ${i\in\N}$,
mutually different with $(K_i)^\perp \neq \emptyset$ such that
$\bigcup_{i=1}^{\infty} K_i=M$.

\textbf{Example 2:}
Any conformal class of a Lorentzian metric,
which is globally hyperbolic without any singularities
determines such a causal structure.
In this case the compact open double cones
form a basis of the usual Euclidean $d+1$ topology.
Each open compact double cone $\sK$ is conformally
equivalent to a copy of Minkowski space.
Consider a spatial Cauchy section $\Sigma$ of $M$
and  a  geodesic world line $p:\tau\to M$ intersecting
$\Sigma$ at $p(0)$, where $\tau$ is the proper time
of the observer.
Now for any $\tau>0$ the causal past of $p(\tau)$
intersects $\Sigma$ in an open set $O_{\tau}$.
Then these open sets  are totally ordered
by their nested inclusion in $\Sigma$,
and their order agrees also with the total
order  of  worldline proper time,
\beq{PO}
O_{\tau_1}\subset O_{\tau_2} \Iff \tau_1<\tau_2 .
\eeq

This is the motivation to consider the
partial order related to the flow of time
and the one related to enlargement in space
to be essentially the same, such that
in the absence of an a priori notion of
a metric time, the nested spatial inclusion
will provide a partial order substituting
time. (Of course this is in essence
similar to the old idea in cosmology of time
given by the volume of a closed, expanding
universe.)

%$\scri^+$

%%%%%%%%%%%%%%%%%%%%%%%%%%%%%%%%%%%%%%%%%%%%%%%%%%%%%%%%%%%%%%%%%%%%%%%%%%
%\section{Causal diffeomorphisms} %3
%\setcounter{equation}{0}
%%%%%%%%%%%%%%%%%%%%%%%%%%%%%%%%%%%%%%%%%%%%%%%%%%%%%%%%%%%%%%%%%%%%%%%%%
Consider now a double cone $\sK$ in $M$
with $O:=\sK\cap\Sigma$ and $\partl O=i^0(\sK)$
and a diffeomorphism $\phi$ in $M$ with pullbacks
$\phi^\Sigma\in\Diff(\Sigma)$ to $\Sigma$
and
$\phi^{\sK}\in\Diff(\sK)$ to $\sK$.
If $\phi(\sK)=\sK$, it can be naturally identified
with an element of $\Diff(\sK)$.
($\phi=id_{M\setminus \sK}$ is a sufficient
but not necessary condition for that
to be true.)
If in addition  $\phi(\Sigma)=\Sigma$
then also $\phi(O)=O$, and $\phi|_O$ is a diffeomorphism of $O$.

Let us now consider a $1$-parameter set of double cones $\{\sK_p\}$
sharing $2$ common null curve segments $n_{\pm}\in\partial\sK_p$
from $i^{\pm}(\sK_p)$ respectively to $i\in i^0(\sK_p)$  which they intersect transversally in $\Sigma$.
Let such cones be parametrized by a line $c$ in $\Sigma$ starting (transversally to  $n$)
at $i$ to some endpoint $f$ on $\partial\Sigma$
(at spatial infinity) such that $p$ is an interior point of $O_p:=\sK_p\cap\Sigma$.
Then we call the limit $W(n_\pm,c):=\lim_{p\to f} K_p$ the wedge
in the surface through $n_\pm$ and $c$.
Note that in the usual (say Minkowski) metric case a wedge has a quite
rigid structure, because $c$ has a canonical location in a surface spanned by
$n_\pm$. The present diffeomorphism invariant analogue is of course much less unique in structure.
%%%%%%%%%%%%%%%%%%%%%%%%%%%%%%%%%%%%%%%%%%%%%%%%%%%%%%%%%%%%%%%%%%%%%%%%%%
\section{Axioms for QFT on  manifolds} %2
%\setcounter{equation}{0}
%%%%%%%%%%%%%%%%%%%%%%%%%%%%%%%%%%%%%%%%%%%%%%%%%%%%%%%%%%%%%%%%%%%%%%%%%%
Clearly QFT on a globally hyperbolic space-time manifold
satisfies isotony (N1), covariance (N2), causality (C), additivity (A)
and existence of a (state dependent GNS) vacuum vector (V).
More particular on Minkowski space there
is is a  unique  Poincare-invariant state $\omega$
such that there is a translational
subgroup of isometries with spectrum in the
closure of the forward light cone only.
However there is no reason
to expect such features
in a more general context.
However, an invariant GNS vacuum vector $\Omega$ still exists
for a  globally hyperbolic space-time,
although in general it depends on the choice
of the state $\omega$.
Hence we will now generalize the axioms of AQFT
from globally hyperbolic space-times to differentiable manifolds.

For a given QFT on manifolds, say the example of quantum gravity
examined below, it remains to check which of the
generalized axioms will hold true.

%%%%%%%%%%%%%%%%%%%%%%%%%%%%%%%%%%%%%%%%%%%%%%%%%%%%%%%%%%%%%%%%%%%%%%%%%%
\subsection{General axioms for QFT on a differentiable  manifold} %2
\setcounter{equation}{0}
%%%%%%%%%%%%%%%%%%%%%%%%%%%%%%%%%%%%%%%%%%%%%%%%%%%%%%%%%%%%%%%%%%%%%%%%%%
%Let CAT be any category of manifolds which is at least $\CO$, i.e. topological
On a differentiable manifold $M$
part of the AQFT structure
can be related to the topological structure of $M$.
The following AQFT axioms are purely topological and
 should hold on arbitrary differentiable manifolds.
Let $M$ be a differentiable manifold  with additional structure $s$ (which may be empty)
and  $\Diff(M,s)$ denote
all diffeomorphisms which preserve $s$.
A $\Diff(M,s)$-invariant algebraic QFT (in the state $\omega$)
can be formulated in terms of axioms on a
net of $*$-algebras ${\cal A}({\cal O})$
(together with a state $\omega$ thereon).
It should at least satisfy
the following axioms:

\nl%\noindent
{\bf N1 (Isotony):}
\bear{N1}
{\cal O}_1\subset {\cal O}_2 & \imp &
{\cal A}({\cal O}_1)\subset {\cal A}({\cal O}_2)\
\forall {\cal O}_{1,2}\in \Diff(M,s)
\ear
\nl%\noindent
{\bf N2 (Covariance):}
\bear{N2}
\Diff(M,s)\ni g {\stackrel{\exists}{\mapsto}} U(g) \in U(\Diff(M,s))&:&
%\nn\\
{\cal A}(g{\cal O})=U(g){\cal A}({\cal O})U(g)^{-1} \ .
\ear
(N1) and (N2) are purely topological,
involving only the mere definition of the net.
These axioms make sense even without a causal structure
(see also \cite{Ra97}).

If ${\cal A}({\cal O})$ is a $C^*$-algebra with norm $||\cdot||$,
it makes sense to impose the following additional axioms:
\nl
{\bf A (additivity):}
\bear{A}
{\cal O} = \cup_{j} {\cal O}_j &\imp &
{\cal A}({\cal O}) = \cl_{||\cdot||}\left(\cup_{j} {\cal A}({\cal O}_j)\right) \ .
\ear
\nl
{\bf V (Invariant Vacuum Vector):}
Given a state $\omega$, there exists
a representation $\pi_\omega$
on a Hilbert space ${\cal H}_\omega$
such that
\bear{V}
\exists \Omega\in{\cal H}_\omega, ||\Omega||=1 &:&
\nn\\
\mbox{\rm (cyclic)}&&
\left(\cup_{\cal O} {\cal R}({\cal O})\right) \Omega
{\stackrel{\rm dense}{\subset}} {\cal H}_\omega
\nn\\
\mbox{\rm (invariant)}&&
U(g)\Omega=\Omega
\ , \quad
g\in \Diff(M,s)\ .
\ear
Note: For any $*$-algebra, the representation $\pi_\omega$ is given by
the GNS construction, ${\cal H}_\omega$ is the GNS Hilbert space.
Properties of $\Omega$ are induced by corresponding properties
of the state $\omega$.
The main issue to check is the invariance under a
unitary representation $U$ of \Diff(M,s).

%%%%%%%%%%%%%%%%%%%%%%%%%%%%%%%%%%%%%%%%%%%%%%%%%%%%%%%%%%%%%%%%%%%%%%%%%%
\subsection{Axioms for QFT on a manifold with cone causality} %2
\setcounter{equation}{0}
%%%%%%%%%%%%%%%%%%%%%%%%%%%%%%%%%%%%%%%%%%%%%%%%%%%%%%%%%%%%%%%%%%%%%%%%%%
With a notion of causality on a differentiable manifold $M$
as defined in the previous section,
the algebraic structure of a QFT
should be related to the causal differential structure of $M$
by further axioms abstracted from
the space-time case.
In this case it is natural to consider nets of
von Neumann algebras.
On a causal differential manifold $M$ (in the sense defined above)
the algebraic structure of a QFT
should satisfy the following axioms which
require the notion of a causal complement.
Let $M$ be a causal differentiable manifold
with additional structure $s$ (which may be empty)
and  $\Diff(M,s)$ denote
all differentiable diffeomorphisms which preserve $s$,
where $s$ is at least a causal structure,
eventually with some additional structure $s'$.
A $\Diff(M,s)$-invariant algebraic QFT in the state $\omega$
is a net of von Neumann-algebras
${\cal R}({\cal O})$
with a state $\omega$
satisfying the following axioms:
\nl
{\bf C (causality):}
\bear{C}
{\cal O}_1\perp {\cal O}_2 &\imp&
{\cal R}({\cal O}_1)\subset {\cal R}({\cal O}_2)' \ .
\ear
\nl%\noindent
{\bf CA (causal additivity):}
\bear{CA}
{\cal O} = \cup_{j} {\cal O}_j &\imp &
{\cal R}({\cal O}) = \left(\cup_{j} {\cal R}({\cal O}_j)\right)'' \ .
\ear
Remarks:
In the case that the net has
both inner and exterior boundary,
\rf{C} had been  weekened in \cite{Ra97}
to a generalization of Haag duality on the boundary of the net.
Here we do not assume a priori the existence of such a
boundary of the net. However an example
of quantum geometry with such a boundary structure
is discussed below.

Given  a net of $C^*$ algebras consistent
with a norm $||\cdot||$, it made sense
to impose (A) above.
If the algebras are in particular also von Neumann ones
(A) should be sharpened to (CA).
In the general case of $*$-algebras (not necessarily $C^*$ ones)
the algebraic closure has no natural topological analogue,
and hence there is
no obvious definition of additivity.
Therefore in \cite{Ra97} neither (A) nor (CA) was assumed.

%%%%%%%%%%%%%%%%%%%%%%%%%%%%%%%%%%%%%%%%%%%%%%%%%%%%%%%%%%%%%%%%%%%%%%%%%
\section{GNS and modular construction} %3
\setcounter{equation}{0}
%%%%%%%%%%%%%%%%%%%%%%%%%%%%%%%%%%%%%%%%%%%%%%%%%%%%%%%%%%%%%%%%%%%%%%%%%
Given a differentiable manifold $M$,
a collection $\{{\cal A}({\cal O})\}_{{\cal O}\in M}$
of $*$-algebras  ${\cal A}({\cal O})$
on bounded open sets ${\cal O} \in M$
is called a {\em net of $*$-algebras},
iff isotony ({\bf N1}) and convariance ({\bf N1}) hold.
%\beq{iso}
%{\cal O}_1\subset {\cal O}_2\imp
%{\cal A}({\cal O}_1)\subset {\cal A}({\cal O}_2)\ .
%\eeq
The net is sometimes also denoted by ${\cal A}^{}_{}:=
{\bigcup_{\cal O}} {\cal A}^{}_{}({\cal O})$.
Selfadjoint elements of ${\cal A}({\cal O})$
may be interpreted as possible measurements in $\cal O$.
Two sets ${\cal O}_1$ and ${\cal O}_2$, related
by a topological isomorphism (e.g. a diffeomorphism) $\chi$
such that $\chi({\cal O}_1)={\cal O}_2$,
may be identified straightforwardly only if there are
no further obstructing relations between them. A relation
like ${\cal O}_1\subset{\cal O}_2$, in addition to the previous one,
implies that
${\cal O}_1$ and ${\cal O}_2$ have to be considered as
non-identical but topologically isomorphic sets.
A similar situation holds on the level of algebras.
Isotony ({\bf N1}) in connection with covariance ({\bf N2})
implies that ${\cal A}({\cal O}_1)$ and ${\cal A}({\cal O}_2)$
are isomorphic but non-identical algebras.

The state of a physical system is mathematical described by a
positive linear functional  ${\omega}$ on $\cal A$.
Given the state ${\omega}$, one
gets via the GNS construction a representation $\pi^{\omega}$ of $\cal A$
by a net of operator algebras on a Hilbert space ${\cal H}^{\omega}$ with
a cyclic vector $\Omega^{\omega}\in {\cal H}^{\omega}$. The
GNS representation $(\pi^{\omega}, {\cal H}^{\omega}, \Omega^{\omega})$
of any state $\omega$ has an associated folium ${\cal F}^{\omega}$,
given as the family of those states $\omega_\rho:=\tr\rho\pi^{\omega}$
which are defined by positive trace class
operators $\rho$ on ${\cal H}^{\omega}$.

Once a physical state $\omega$
(which implicitly contains all peculiarities of a particular
observation)
has been specified,
one can consider in each
algebra ${\cal A}({\cal O})$ the equivalence relation
\begin{equation}
A\sim B \ \ :\Iff \ \
{\omega}^{\prime}(A-B)=0,\ \ \forall  {\omega}^{\prime}\in{\cal F}^{\omega}\ .
\end{equation}
These equivalence relations generate a two-sided ideal
\beq{ideal}
{\cal I}^{\omega}({\cal O}):=\{A\in{\cal A}({\cal O})\vert
{\omega}^{\prime}(A)=0\}
\eeq
in ${\cal A}({\cal O})$.
The (dynamically relevant) state dependent algebra of { observables}
${\cal A}^{\omega}_{}({\cal O}):=\pi^{\omega}({\cal A}({\cal O}))$ may
be constructed from
the (kinematically relevant) algebra of
observation procedures ${\cal A}({\cal O})$ by taking
the quotient
\begin{equation}
{\cal A}^{\omega}_{}({\cal O}) =
{\cal A}({\cal O})/{\cal I}^{\omega}({\cal O})\ .
\end{equation}
The net of state-dependent algebras then is also denoted as
${\cal A}^{\omega}_{}:=
{\bigcup_{\cal O}} {\cal A}^{\omega}_{}({\cal O})$.
By construction, any diffeomorphism $\chi\in \Diff(M)$ induces an
algebraic isomorphism $\alpha_{\chi}$ of the observation procedures.
Nevertheless, for a given state $\omega$,
the action of $\alpha_{\chi}$ will
in general {\em not} leave ${\cal A}^{\omega}_{}$
invariant.
In order to satisfy
\begin{equation}
\label{D1}
\alpha_{\chi}({\cal A}^{\omega}_{}({\cal O}))=
{\cal A}^{\omega}_{}(\chi({\cal O}))\ .
\end{equation}
the ideal
${\cal I}^{\omega}({\cal O})$ must transform covariantly, i.e.
the diffeomorphism ${\chi}$ must satisfy
the condition
\begin{equation}
\label{D2}
\alpha_{\chi}({\cal I}^{\omega}({\cal O}))=
{\cal I}^{\omega}(\chi({\cal O}))\
\end{equation}
for some algebraic isomorphism $\alpha_{\chi}$.
Due to non-trivial constraints \rf{D2}, the (dynamical) algebra of
observables, constructed with respect to
the folium ${\cal F}^\omega$, does in general no longer exhibit
the full $\Diff(M)$ symmetry of the (kinematical) observation procedures.
The symmetry of the observables is dependent on (the folium of) the state
$\omega$.
Therefore, the selection of a folium of states
${\cal F}^\omega$, induced by the actual choice of a state $\omega$,
results immediately in a breaking of the $\Diff(M)$ symmetry.
The diffeomorphisms which satisfy
the constraint condition (\ref{D2}) form a subgroup called
the {\em dynamical group} of the state $\omega$.
$\alpha_{\chi}$ is called a {\em dynamical} isomorphism
(w.r.t. the given state $\omega$) w.r.t. $\chi$, if (\ref{D2})
is satisfied.

The remaining dynamical symmetry group, depending
on the folium ${\cal F}^\omega$ of states related to $\omega$,
has two main aspects which we have to examine
in order to specify the physically admissible states:
Firstly, it is necessary to specify its state dependent
algebraic action on the net of observables. Secondly, one has to
find a geometric interpretation for the group and its action on $M$.

If we consider the dynamical group as an {\em inertial},
and therefore global, manifestation of dynamically ascertainable
properties of observables,
then its (local) action should be correlated with (global)
operations on the whole net of observables.
This implies that at least some of the dynamical
isomorphisms $\alpha_\chi$ are not inner.
(For the case of causal nets of algebras
it was actually already shown
that, under some additional assumptions,
the isomorphisms of the algebras are in general not inner\cite{Wo}.)

In the following we consider  instead of the net of observables
${\cal A}^{\omega}_{}({\cal O})$
the net of associated von Neumann algebras
${\cal R}^{\omega}_{}({\cal O})$.

Let us assume
$\Omega^{\omega}$ is a cyclic and separating vector for
${\cal R}^{\omega}_{\min}$ and ${\cal R}^{\omega}_{\max}$.
By isotony it is so for any
local von Neumann algebra ${\cal R}^{\omega}_{}({\cal O}^{x}_{s})$
too.
As a further consequence, on any region ${\cal O}^{x}_{s}$,
the Tomita operator $S$ and  its conjugate $F$
can be defined densely by
\begin{equation}
\label{T0}
S A \Omega^{\omega}:= A^{*} \Omega^{\omega} \ \ \mbox{for}\ \ A\in
{\cal R}^{\omega}_{}({\cal O}^{x}_{s})\ ,
\end{equation}
\begin{equation}
\label{T1}
F B \Omega^{\omega}:=B^{*} \Omega^{\omega}
 \ \ \mbox{for}\ \ B\in
{{\cal R}^{\omega}_{}({\cal O}^{x}_{s})}' \ .
\end{equation}
The  closed Tomita operator $S$ has a polar decomposition
\begin{equation}
S=J\Delta^{1/2} \ ,
\label{T2}
\end{equation}
where  $J$ is antiunitary and $\Delta:=FS$ is the self-adjoint, positive
modular operator.
The Tomita-Takesaki theorem \cite{Ha} provides us with a one-parameter
group of state dependent isomorphisms $\alpha^{\omega}_t$ on
${\cal R}^{\omega}_{}({\cal O}^{x}_{s})$,
defined by
\begin{equation}
\alpha^{\omega}_t (A)= \Delta^{-it}\ A\ \Delta^{it}\ ,   \ \ \mbox{for}\ \
A\in{\cal R}^{\omega}_{\max} \ .
\label{T3}
\end{equation}
A strongly continuous unitary
implementation  of the modular group of
the $1$-parameter family of isomorphisms (\ref{T3})
w.r.t. $\omega$
is given by conjugate action of operators
$e^{-it\ln\Delta}$, ${t\in\R}$.
By (\ref{T3}) the modular group, for a state $\omega$
on the net of von Neumann algebras,
defined by ${\cal R}^{\omega}_{\max}$,
might be considered as a $1$-parameter subgroup of the dynamical group.
Note that, with Eq. (\ref{T1}), in general,
the modular operator $\Delta$ is not located on
${\cal O}^{x}_{s}$. Therefore,  in general, the modular isomorphisms
(\ref{T3}) are not inner.
The modular isomorphisms are known to act as
{inner} isomorphisms, iff the von Neumann algebra
${\cal R}^{\omega}_{}({\cal O}^{x}_{s})$
generated by $\omega$
contains only semifinite factors (type I and II), i.e.
$\omega$ is a semifinite trace.

Above we considered concrete von Neumann algebras
${\cal R}^{\omega}_{}({\cal O}^{x}_{s})$, which are in fact
operator representations of an abstract von Neumann algebra
${\cal R}$ on a GNS Hilbert space ${\cal H}^{\omega}$ w.r.t.
a faithful normal state $\omega$.
In general, different faithful normal states generate different
concrete von Neumann algebras and different modular isomorphism groups
of the same abstract  von Neumann algebra.

The outer modular isomorphisms form the cohomology group
$\Out {\cal R}:=\Aut {\cal R}/\Inn {\cal R}$
of modular isomorphisms modulo inner modular isomorphisms.
This group is characteristic
for the types of factors contained in the von Neumann algebra\cite{Co}.
Per definition $\Out {\cal R}$ is trivial for inner isomorphisms.
Factors of type III${}_1$ yield $\Out {\cal R}=\R$.
This is a physically well-known case, realized
by standard QFT, say on Minkowski space.
Here a distinguished
$1$-parameter group of outer modular isomorphisms
is given, which may be interpreted geometrically as a certain subgroup
of the dynamical group.
%%%%%%%%%%%%%%%%%%%%%%%%%%%%%%%%%%%%%%%%%%%%%%%%%%%%%%%%%%%%%%%%%%%%%%%%%%
\section{Application for quantum gravity} %6
\setcounter{equation}{0}
%%%%%%%%%%%%%%%%%%%%%%%%%%%%%%%%%%%%%%%%%%%%%%%%%%%%%%%%%%%%%%%%%%%%%%%%%%
On a region causally exterior
to a topological horizon $\sH$,
on any $d$-dimensional spatial slice $\Sigma$,
there exists a net of Weyl algebras
for states with
an {\em infinite} number of intersection
points of edges and
transversal $(d-1)$-faces
within any neighbourhood of the spatial boundary
$\sH{\rm }\cap\Sigma\iso S^2$.

$\Sigma$ be a spatial slice.
C-causality constrains the algebras
localized within $\Sigma$.
On $\Sigma$ it should hold
\begin{equation}\label{discom}
{\cal O}_1\cap {\cal O}_2=\emptyset
\quad \imp \quad [{\cal A}({\cal O}_1), {\cal A}({\cal O}_2)]=0 .
\end{equation}

A (spin network) state $\omega$ over the algebra ${\cal A}(\Sigma)$
may be defined by a closed, oriented differentiable graph $\gamma$
 embedded in $\Sigma$,
with an infinite  number of differentiable edges $e\in E$
intersected transversally by a
differentiable $d-1$-dimensional oriented surface $S$
at a countable of intersection vertices $v\in V$.
Let $C_{\gamma}\in \Cyl$ be a $\Cinf$ Cylinder function with respect to an
$\SO(d)$ holonomy group  on $\gamma$, i.e.
on each closed {\em finite} subgraph $\gamma'\subset\gamma$
it is $C_{\gamma'}:=c(g_1,\ldots,g_N)$ where $g_k\in\SO(d)$,
and $c$ is a differentiable function.
%%%%%
%%%%%
With test function $f$
the action of a derivation $X_{S,f}$ on $\Cyl$
 is defined by
\beq{deriv}
 X_{S,f}\cdot C_{\gamma}:=\frac{1}{2}
 \sum_{v\in V}\sum_{e_v\in E:\partl e_v\ni v}
 \kappa(e_v) f^i(v) X^i_{e_v} \cdot c ,
\eeq
where $\kappa(e_v)=\pm1$ above/below $S$
(for the following purposes we may just exclude the tangential
case $\kappa(e_v)=0$)
and $ X^i_{e_v} \cdot c $ is the action of the left/right invariant
vector field (i.e. $e_v$ is oriented away from/towards the
surface $S$) on the argument of $c$ which corresponds to the
edge $e_v$.
%%%%%
Let $\Der$ denote the span of all such derivations.

Here the classical (extended) phase space is the cotangent
bundle  $\Gamma=T^*{\mathcal{C}}$
over a space $\mathcal{A}$ of (suitably regular) finitely localized
connections.
Let $\delta = ( {\delta_A} , {\delta_E} ) \in T_{e}\Gamma$.
With suitable boundary conditions,
a (weakly non-degenerate) symplectic form $\Omega$ over $\Gamma$
acts via
\bear{sympl}
\Omega|_{(A_e,E_e)}
\left ( \delta, \delta' \right )
&:=&
{1 \over \ell^2} \int_{\Sigma} {\rm Tr}
\left [  *E \wedge A'  -   *E' \wedge A  \right ] .
\ear
After lifting from $\mathcal{C}$ to $\Gamma$,
the cylinder functions $q\in\Cyl$ serve as
(gauge invariant) classical elementary
configuration functions on $\Gamma$.
The derivations $p\in\Der$ serve as classical elementary momentum
functions on $\Gamma$. They are obtained as the Poisson-Lie action
of $2$-dimensionally smeared duals of densitized triads $E$.
$\Cyl\times\Der$ has a Poisson-Lie structure
\beq{Poisson}
\{(q,p),(q',p'))\}:=(p q'- p' q  ,[p,p']) ,
\eeq
where $[p,p']$ denotes the Lie bracket of $p$ and $p'$.
An antisymmetric bilinear form on $\Cyl\times\Der$ is given by
\bear{pres}
\Omega_{0}
\left ( {(\delta_q,\delta_p)}, {(\delta_q',\delta_p')} \right )
&:=&
\int_{{\mathcal{C}_{\gamma\cup\gamma'}}/
{\mathcal{G}_{\gamma\cup\gamma'}}}d\mu_{\gamma\cup\gamma'}
\left [  p  q'  -   p' q  \right ] ,
\ear
where $q,q'\in \Cyl$ have support on $\gamma$ resp. $\gamma'$,
with $p  q' -  p' q\in \Cyl$
integrable over ${\mathcal{C}_{\gamma\cup\gamma'}}/
{\mathcal{G}_{\gamma\cup\gamma'}}$ with
measure $d\mu_{\gamma\cup\gamma'}$.

On $T_e\Gamma$, the symplectic form $\Omega$  yields functions
of the form $\Omega((\delta_A,\delta_E),\cdot)$.
Canonical quantization then associates
to any function $\Omega(f,\cdot)$
a selfadjoint operator $\hat\Omega(f,\cdot)$
and a corresponding
unitary Weyl element $W(f):=e^{i\hat\Omega(f,\cdot)}$,
both on some extended Hilbert space.
With multiplication
$W(f_1) W(f_2) := e^{i\Omega(f_1,f_2)}W(f_1+f_2)$,
and conjugation $*: W(f)\mapsto W(-f)$,
the Weyl elements generate a $*$-algebra.
A norm on $\Gamma$ is defined by
$\|f\|:=\frac{1}{4}\sup_{g\neq 0}\frac{\Omega(f,g)}{<g,g>}$.
The $C^*$-closure under the sup-norm then generates
a  $C^*$-algebra $CCR(W(f),\Omega)$.
With regular $\Omega$ this CCR Weyl algebra is simple,
i.e. there is no ideal.
Observables of quantum $3$-geometry are
then the selfadjoint elements within a
gauge and $3$-diffeomorphism invariant $C^*$-subalgebra
${\cal A}_{\gamma}\subset C^*(W(f),f\in\Gamma)$.
In a gauge and $3$-diffeomorphism invariant
representation of ${\cal A}_{\gamma}$,
typical observables in are represented by
configuration multiplication operators $C_{\gamma}\in \Cyl$
on Hilbert space $\mathcal{H}_{\gamma}$,
and by gauge-invariant and 3-diffeomorphism invariant
combinations of derivative operators $X_{S,f}\in \Der$,
like e.g. a certain quadratic combination which yields
the area operator.
%Note however that for loop quantum gravity $\Omega$ need a priori not
%be simple. There may be free loops in $\gamma$ which do not intersect
%given surfaces $S$. These free loops form an Abelian ideal $J_{\gamma}$ in $A_{\gamma}$,
%whence $A_{\gamma}$ is not simple.
%
%However in the case of a degenerate symplectic form, passing
%to the quotient  $A_{\gamma}/J_{\gamma} $
%yields again a simple algebra.
%For loop quantum gravity the phase space is spanned by
%generalized functions
%$f=(\mathcal{C}_{\gamma},T_{S,\cdot})\in \Cyl\times\Der$.

For each finite $\gamma'\subset\gamma$, the sets $E(\gamma')$ and $V(\gamma')$
of edges resp. vertices of $\gamma'$ are finite. Then the
connections
${\mathcal{C}}_{\gamma'}=\prod_{e\in E(\gamma')}G_e\cong G^{E(\gamma')}$
and the gauge group
${\mathcal{G}}_{\gamma'}=\prod_{v\in V(\gamma')}G_v\cong G^{V(\gamma')}$
on $\gamma'$ inherit a unique measure from the measure on $G$
(for compact $G$ the Haar measure).
The action of $\mathcal{G}_{\gamma'}$ on $\mathcal{C}_{\gamma'}$ is defined by
$(g A)_e := g_{t(e)} A_e g_{s(e)}^{-1}$ where $s$ and $t$ are the source and target
functions $E(\gamma')\to V(\gamma')$ respectively.
This action gives rise to gauge orbits and a corresponding projection
$\mathcal{C}_{\gamma'}\twoheadrightarrow \mathcal{C}_{\gamma'}/\mathcal{G}_{\gamma'}$.
The projection induces the measure on $C_{\gamma'}/\mathcal{G}_{\gamma'}$.
Bounded functions w.r.t. to that measure define then
the gauge invariant Hilbert space
$\mathcal{H}_{\gamma'}:=L(\mathcal{C}_{\gamma'}/\mathcal{G}_{\gamma'})$.
%%% end corrected 2000

However, over finite graphs, all is still QM rather than QFT.
In order to obtain
an infinite number of degrees of freedom
on any finite localization domain which includes
the inner boundary $S^{d-1}$ (the intersection $S^{d-1}$
of $\sH$ and $\Sigma$),
let  $S^{d-1}$ be intersected
by an infinite number of edges of some graph $\gamma$
in the exterior spatial neighborhood of  $\sH$.
In the $3+1$-dimensional case, evaluation of the area operator
on the puncture of the boundary $S^2$ from edge $p$
yields a quantum of area proportional to
$j_p(j_p+1)$ for edge $p$ carrying a spin-$j_p$
representation of the group $G$.
Since $S^2$ is compact the punctures should have at least one
accumulation point.
Hence for typical configurations in the principal representation,
near that accumulation point the area will explode to infinity.
When almost all punctures are located in arbitrary small neighborhoods
of a finite number $n$ of accumulation points,
corresponding states represent quantum geometries of a black hole with $n$
stringy hairs extending out to infinity. In particular, the $n=1$ case was discussed
in more detail in \cite{MR99bh}.
\section*{Acknowledgment} %6
%\setcounter{equation}{0}
%%%%%%%%%%%%%%%%%%%%%%%%%%%%%%%%%%%%%%%%%%%%%%%%%%%%%%%%%%%%%%%%%%%%%%%%%%
I would like to thank for the hospitality of the
Instituto de Matematicas y Fisica Fundamental
at CSIC Madrid, where part of the manuscript was completed.
%This condition is e.g. also
%satisfied for Borchers algebras.
%Of course the inverse of (\ref{discom})
%is not true in general.

\np
%\section*{References}

\end{document}